\documentclass{article}
\usepackage{opex3}
\usepackage{color}
\usepackage{upgreek}
\usepackage{amsmath}
\usepackage{amssymb}
\usepackage{graphicx}
\usepackage{epstopdf}

\begin{document}

\title{Ideal type-II Weyl points in twisted one-dimensional dielectric photonic crystals }

\author{Ying Chen$^{1,2,*}$, Hai-xiao Wang$^3$,Qiaoliang Bao$^4$, Jian-Hua Jiang$^5$, Huanyang Chen$^{2,\dag}$} 
	

\address{\noindent{$^1$College of Information Science and Engineering, Fujian Provincial Key Laboratory of Light Propagation and Transformation, Huaqiao University, Xiamen 361021, China.}\\
$^2$Department of Physics and Institute of Electromagnetics of Acoustics, Xiamen University, Xiamen 361005, China.\\
$^3$School of Physical Science and Technology, Guangxi Normal University, Guilin 541004, China.\\
$^4$Department of Applied Physics, the Hong Kong Polytechnical University, Hong Kong, China.\\
$^5$School of Physical Science and Technology, and Collaborative Innovation Center of Suzhou Nano Science and Technology, Soochow University, 1 Shizi Street, Suzhou 215006, China.}

\email{$^*$yingchen@hqu.edu.cn ; $^\dag$kenyon@xmu.edu.cn }



\begin{abstract}
	
Weyl points are the degenerate points in three-dimensional momentum space with nontrivial topological phase, which are usually realized in classical system with structure and symmetry designs. Here we proposed a one-dimensional layer-stacked photonic crystal using anisotropic materials to realize ideal type-II Weyl points without structure designs. The topological transition from two Dirac points to four Weyl points can be clearly observed by tuning the twist angle between layers. Besides, on the interface between the photonic type-II Weyl material and air, gappless surface states have also been demonstrated in an incomplete bulk bandgap. By breaking parameter symmetry, these ideal type-II Weyl points at the same frequency would transform into the non-ideal ones, and exhibit topological surface states with single group velocity. Our work may provide a new idea for the realization of photonic Weyl points or other semimetal phases by utilizing naturally anisotropic materials.
\end{abstract}

\section{Introduction}
Three-dimensional(3D) topological semimetal phases have attracted much attentions in recent years, such as Weyl points[1-4], Dirac points[5,6] and nodal lines[7,8], while Dirac and Weyl points can further be classified to type-I and type-II points depending on their equi-frequency curves at Dirac/Weyl frequency[9-11]. As the fundamental state, 3D Dirac points can be seen as a pair of Weyl points with opposite charges, and may be separated in momentum space by breaking time-reversal symmetry or inversion symmetry. Specifially, Dirac/Weyl semimetals have been demonstrated to exhibit several peculiar phenomena, such as Fermi arcs and chiral anomaly [1,11], while for type-II Weyl points, the unique features including tilted cone dispersion and the surface states existed in an incomplete bandgap would bring more exotic effects[10]. 

In classical systems, the analogues of various topological semimetals have also been achieved  in photonics[12-25], acoustics[26-31],circuits[32] and magnetic systems[33,34]. Type-I Weyl points and the arc surface states have been theoretically[12] and experimentally[13,14,27] verified in 3D photonic/phononic crystals and metamaterials, and also been found recently that the Weyl semimetals can sustain higher-order topological hinge states[35-38]. With the discovery of type-II Weyl points in condensed matter[10], classical type-II Weyl points have been subsequently realized in a photonic waveguide array[19] and chiral hyperbolic metamaterial[20], together with the topological surface states, and further been achieved in phononic crystals[39,40] and electric circuits[41]. Among them, the ideal Weyl points, which are usually symmetry-related, have received particular interests focusing on achieving degenerate nodal points with the same frequency in various bosonic systems[41-45]. However, it is shown that in bosonic system these Weyl semimetals are usually realized using 3D artificial periodical structures with complicated structure designs, and the ideal type-II Wely points is still beyond realization in photonic system. Besides, for the topological transition from Dirac points to Weyl points in classical semimetals, previous works usually achieved this effect in two artificial structures with different symmetries or couplings [18,31], which is very inflexible and untunable. 

\begin{figure*}[t]
	\centering
	\includegraphics[height=7.5cm]{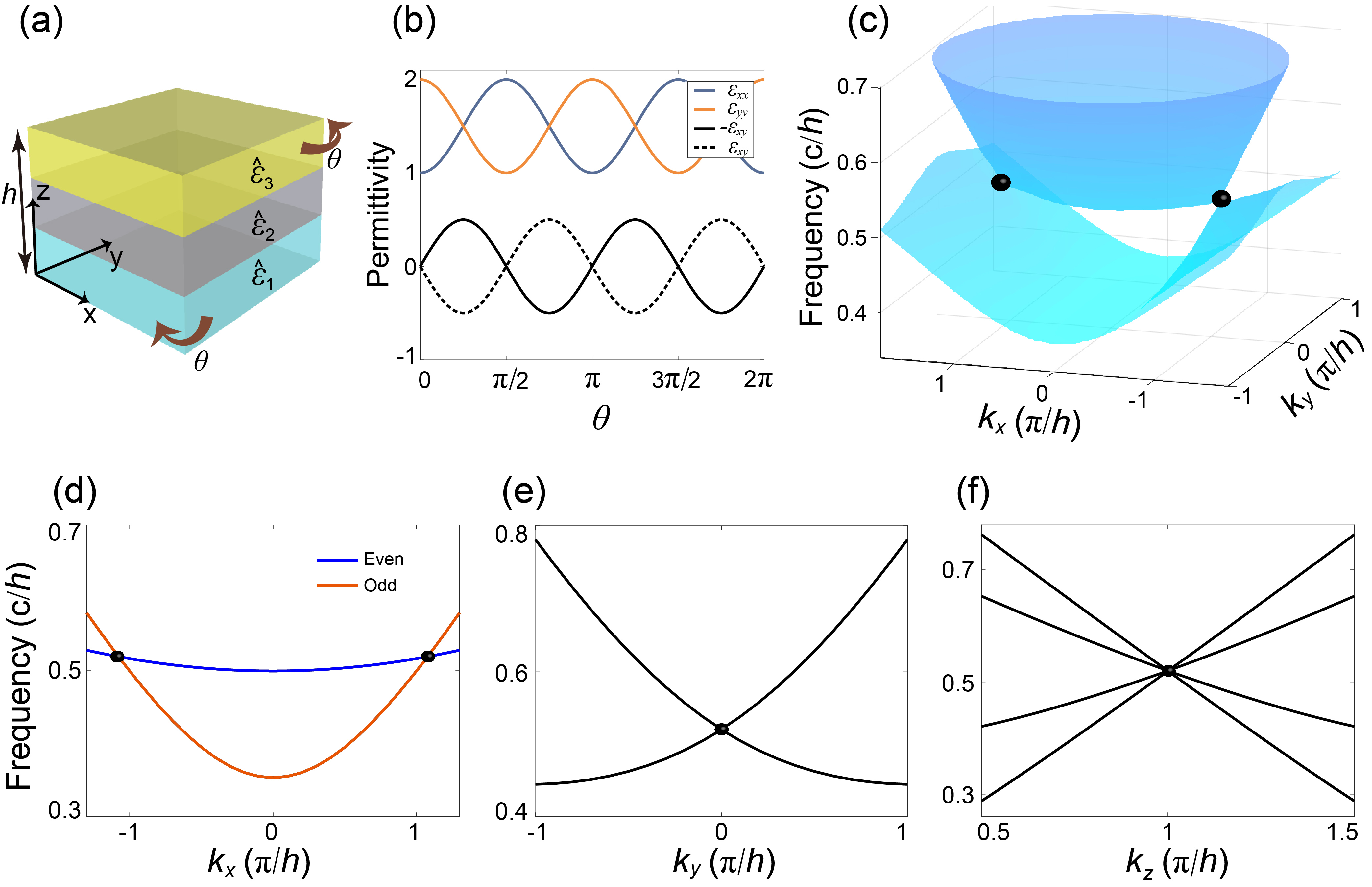}
	\caption{(a) Schematic of the unit cell of the twisted layer-stacked photonic crystals, with relative permittivity tensor $\epsilon_1$, $\epsilon_2$ and $\epsilon_3$ from lower to upper layers, and twist angle $\theta$ for the first(clockwisely) and third(counterclockwisely) dielectrics. The period in $z$ direction is $h$ with three layers equally divided to $h/3$, while in $xy$ plane the dielectrics are continuous.(b) Dependence of the permittivity components on twist angle $\theta$. (c) 3D band structure of the twisted photonic crystal on $k_x$-$k_y$ plane with respect to $k_z=\pi/h$ when $\theta=0^{\circ}$, the pair of black dots show the Dirac points ($k_x=\pm1.07\pi/h$, $k_y=0$) with fourfold degeneracies.(d),(e) and (f) The linear dispersion in three orthogonal directions around the type-II Dirac points (0.52$c/h$).The blue and red lines in (c) indicate the bands with even and odd $M_y$ parities. }
	\label{fg1}
\end{figure*}

In this paper, we will show that the simple layer-stacked one-dimensional(1D) photonic crystal could exhibit ideal type-II Weyl points by introducing the degree of freedom of twist angle. Utilizing nondispersive anisotropic dielectrics combined with twist angles, we are able to realize a topological photonic system without structure design. The evolution from two type-II Dirac points to four type-II Weyl points can be easily achieved by tuning the twist angles, which induces topological phase transitions, and the gapless surface states have also been observed in an incomplete bandgap. Moreover, we further obtain the non-ideal type-II Weyl points and surface states with single group velocity by breaking the parameter symmetry. Our work may pave a new way for the study of Weyl system and the tuning between Dirac and Weyl points.

\section{Ideal type-II Weyl points and surface states in the twisted photonic crystals}

Here we propose a 1D layer-stacked photonic crystal using constant anisotropic materials, with three types of dielectric layers as unit cell (see FIG. 1(a)). The designed photonic crystal is periodic along $z$ direction with a lattice constant $h$, while keeps homogeneous in the $xy$ plane. We set the three layers to have the same thickness of $h/3$, and adopt identical nondispersive anisotropic dielectrics with the relative permittivity $\epsilon_0 = $ diag $ [\epsilon_{11}, \epsilon_{22}, \epsilon_{33}]$, whereas the dielectrics of lower (upper) layer rotate an angle $\theta$ counterclockwisely (clockwisely) from the $x$ axis, while the middle layer keep unchanged. In this way, the relative permittivity tensor of the unit cell of such a photonic crystal can be expressed as
\begin{equation}
\epsilon_1=\begin{bmatrix}
\epsilon_{xx}& \epsilon_{xy} &0 \\
\epsilon_{yx} &\epsilon_{yy}& 0\\
0& 0 &\epsilon_{zz}
\end{bmatrix}; 
\epsilon_2=\epsilon_0;
\epsilon_3=\begin{bmatrix}
\epsilon_{xx}& -\epsilon_{xy} &0 \\
-\epsilon_{yx} &\epsilon_{yy}& 0\\
0& 0 &\epsilon_{zz}
\end{bmatrix}
\end{equation}
where $\epsilon_{xx}=\textup{cos} ^2\theta\epsilon_{11}+\textup{sin}^2\theta\epsilon_{22}$, $\epsilon_{yy}=\textup{sin} ^2\theta\epsilon_{11}+\textup{cos}^2\theta\epsilon_{22}$, $\epsilon_{xy}=\epsilon_{yx}=(\epsilon_{11}-\epsilon_{22})\textup{sin}\theta\textup{cos}\theta$, and $\epsilon_{zz}=\epsilon_{33}$ are all constant for a fixed $\theta$ (see FIG. 1(b) for the dependence of  permittivity components on $\theta$). Thus, this 1D photonic crystal can actually be considered as a three-dimensional(3D) structure due to the degree of freedom of the twist angle $\theta$.

\begin{figure*}[t]
	\centering
	\includegraphics[height=7cm]{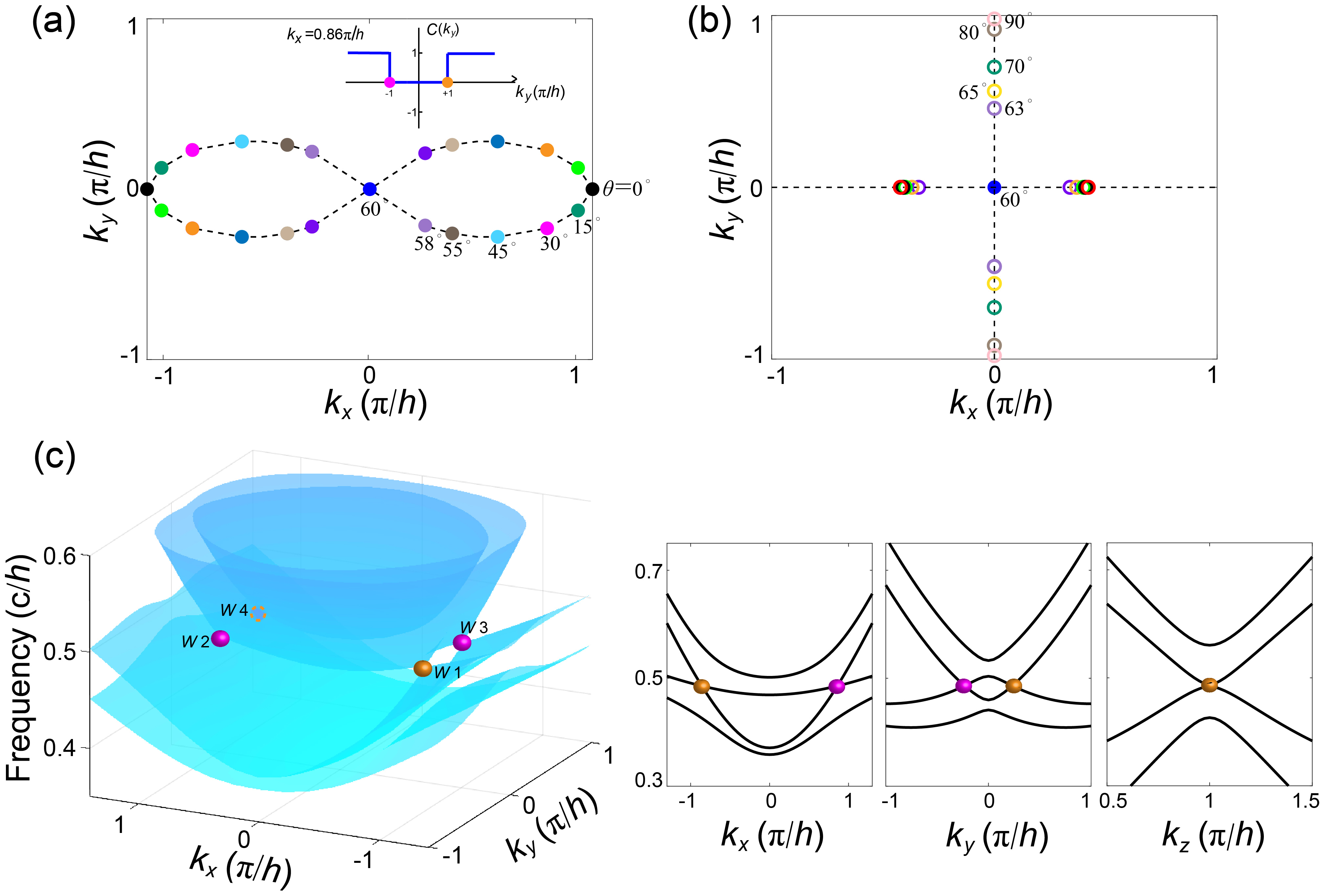}
	\caption{(a) The evolutions of Dirac and Weyl points along with the twist angle $\theta$ from $0^{\circ}$ to $60^{\circ}$ when $k_z=\pi/h$. The inset shows Chern numbers and topological charges as a function of $k_y$ for fixed $k_x$=0.86$\pi/h$, which is related to the Weyl points in (c) for $\theta=30^{\circ}$. (b) The evolutions of Weyl points when $\theta$ changes from $60^{\circ}$ to $90^{\circ}$. (c) Left: 3D band structure of the twisted photonic crystal on $k_x$-$k_y$ plane with $k_z=\pi/h$ when $\theta=30^{\circ}$, the yellow ($W$1 and $W$4) and magenta ($W$2 and $W$3) dots denote Weyl points with +1 and -1 charges. Right: Corresponding dispersions near the Weyl points in three orthogonal directions. Four ideal type-II Weyl points are located at ($\pm0.86$, $\pm0.23$, 1)$\pi/h$ and residing at the same frequency 0.48$c/h$.}
	\label{fg2}
\end{figure*}

Considering the initial case, we first numerically calculate the band structure at $k_z=\pi/h$ plane using COMSOL Multiphysics, with the relative permittivity $\epsilon_{11}=1$, $\epsilon_{22}=2$ and $\epsilon_{33}=14$, while the twist angle $\theta=0^\circ$. In this case, the twisted photonic crystal is reduced to a homogeneous anisotropic material and the dispersions for lowest four bands are presented in FIG. 1(c), which are doubly degenerate for each two bands. Besides, there is a pair of degenerate points with fourfold degeneracy (the black dots) at ($k_x=\pm1.07\pi/h$, $k_y=0$). It should be emphasized that the positions of degenerate points can locate outside 0 to $\pi/h$ due to the non-periodicity in the $xy$ plane. To further demonstrate the properties of these degenerate points, the corresponding 1D band structure in three orthogonal directions are shown in FIG. 1(d)-1(f). We first sweep $k_x$ for a fixed $(k_y, k_z)=(0, 1)\pi/h$  and obtain two degenerate points at $k_x=\pm1.07\pi/h$, as shown in FIG. 1(d). For the intrinsic symmetry of electromagnetism in the anisotropic permittivity[23],the $M_y$-even (with only ($E_x$, $E_z$, $H_y$)) and $M_y$-odd (with only ($H_x$, $H_z$, $E_y$)) states with respect to the $k_y$=0 mirror plane are then found in this photonic crystal when $\theta=0^\circ$, as denoted by the blue and red bands in FIG. 1(d). Thus the two black dots are exactly the crossing of bands with opposite parities, which indicates the Dirac points. For each point, the two tilted bands with the same sign of group velocity may be further classified as a type-II Dirac point. In FIG. 1(e), by fixing $(k_x, k_z)=(1.07, 1)\pi/h$ and sweeping $k_y$, we get a degenerate point at $k_y$=0. Similarly, there is a fourfold degenerate point situated at $k_z=\pi/h$ when we change $k_z$ at $(k_x, k_y)= (1.07, 0)\pi/h$, as shown in FIG. 1(f). Therefore, the tilted linear dispersions around the two degenerate points verify the existence of type-II Dirac points for $\theta=0^\circ$.

\begin{figure*}[tbh]
	\centering
	\includegraphics[height=7cm]{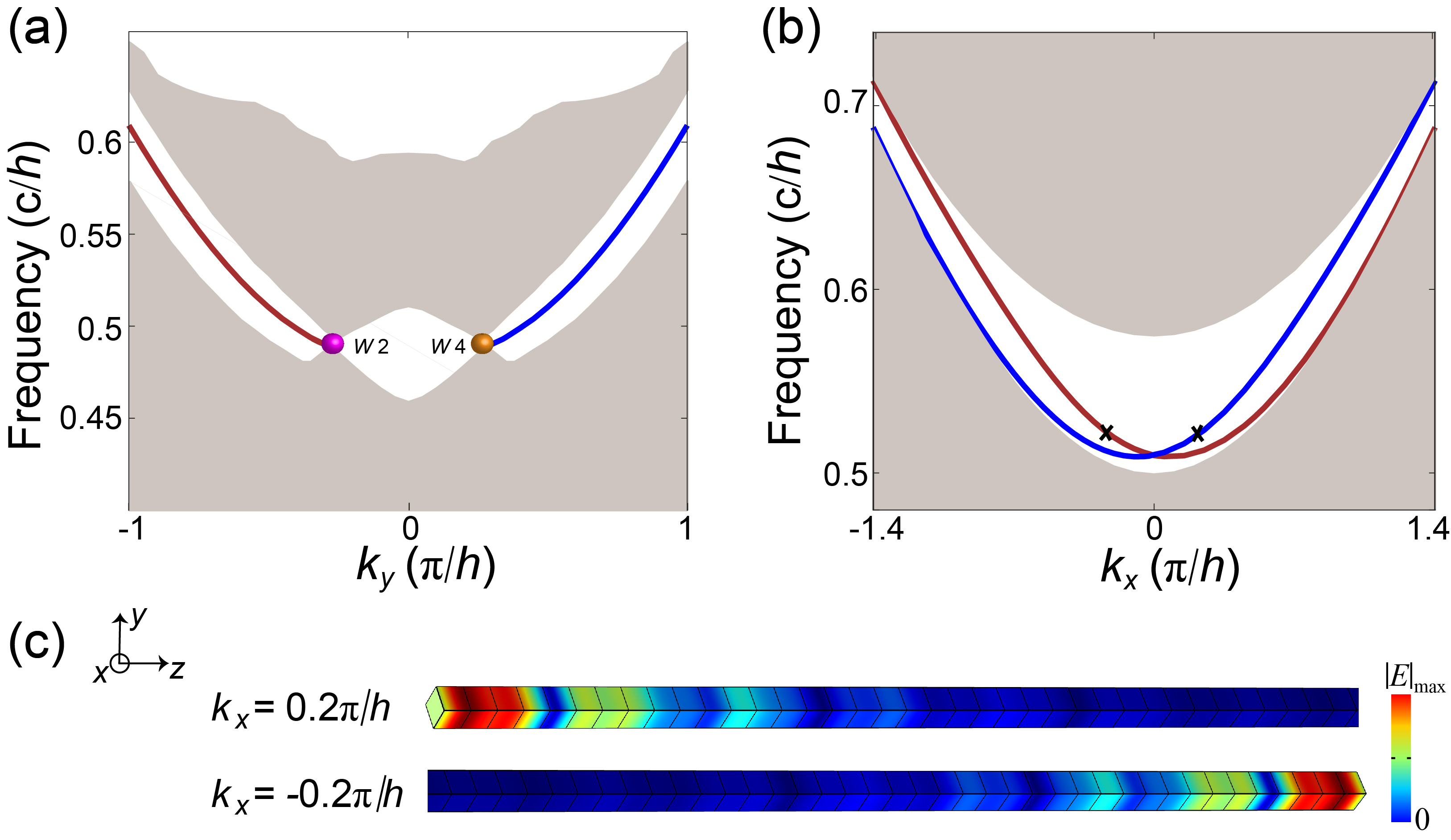}
	\caption{ (a) The projected spectrum plotted as a function of $k_y$ for $k_x$=0.86$\pi/h$ and $\theta$=30$^\circ$, where the supercell is finite along the $z$ direction and infinite in the $x$ and $y$ directions. (b) The projected spectrum for fixed $k_y$=$\pi/h$ and $\theta$=30$^\circ$. Surface states ( bands in red and blue) are existed in an incomplete bandgap, and the bulk bands are shaded in grey. (c) The eigenfields |$\bf{E}$| of the surface states marked by the black crosses at $0.52c/h$ in (b) for $k_x$=-0.2$\pi/h$ and $k_x$=0.2$\pi/h$, respectively.}
	\label{fg3}
\end{figure*}

Next, we change $\theta$ from $0^\circ$ to $90^\circ$ and study the evolution of degenerate points with twist angle $\theta$ at the $k_z=\pi/h$ plane. For the dielectrics in Eq. (1), the period of permittivities $\epsilon_1$ and $\epsilon_3$ is $\theta=180^\circ$ (see FIG. 1(b)), while the band structure for $\theta=\theta_1$ is identical to that for $\theta=(180^\circ-\theta_1)$ when $0<\theta_1<180^\circ$, thus we only display the degeneracies of band structure for $\theta$ ranging from $0^\circ$ to $90^\circ$ for simplicity, as shown in FIG. 2(a) and 2(b). Firstly, we study the evolution of these degenerate points from $0^\circ$ to $60^\circ$ in FIG. 2(a). It is shown that the original two type-II Dirac points (black dots) are split into two pairs of type-II Weyl points for the nonzero twist angle $\theta$, which is formed by the twofold degeneracy of two middle bands (see the 3D band structure in FIG. 2 (c) for $\theta=30^\circ$). Besides, each pair of Weyl points with the same $k_x$ will carry +1 and -1 charges, respectively, indicating the nontrivial topology of these type-II Weyl points. The inset for Chern number and topological charges related to the type-II Weyl points at $k_x=0.86\pi/h$ further verifies this point. During this process, we can see that the $k_x$ coordinates of Weyl points gradually decrease with the increase of $\theta$ and evatually degenerate at $k_x=0$ when $\theta=60^\circ$. While for the $k_y$ coordinates of these Weyl points, they will first undergo an increasing process and then gradually decrease to $k_y=0$ at $\theta=60^\circ$. Thus the evolutionary path of these four Weyl points forms an "eye-glass" shape in momentum space when twist angle $\theta$ changes from $0^\circ$ to $60^\circ$. Specifically, when the photonic crystal has a twist angle of $60^\circ$, its band structure is the same as that for $\theta=120^\circ$, which only presents a degenerate point at $(0, 0, \pi/h)$(denoted as $A$). It can be explained that this layer-stacked crystal has a complete rotation period of $360^\circ$ along $z$ direction when $\theta=120^\circ$ and dispalys a three-fold screw symmetry $\hat{S}_{3z}$. Together with time-reversal symmety $T$, the combined symmetry $T\hat{S}_{3z}$ guarantees that the twofold degeneracy must be at high symmetry $A$ point or $\Gamma$ point [$(0, 0, 0)$] in momentum space[46].

The evolution of these type-II Weyl points for $\theta$ ranging from $60^\circ$ to $90^\circ$ is shown in Fig. 2(b). Starting from a degenerate point at $(0, 0, \pi/h)$ for $\theta=60^\circ$, four type-II Weyl points then appear at the high symmetry lines of $k_x=0$ and $k_y=0$ when twist angle $\theta$ continues to increase. In detail, for a fixed $\theta$ the degenerate points located at $k_x>0$ and $k_y>0$ regions belong to a pair of Weyl points with opposite charges, while the degenerate points at $k_x<0$ and $k_y<0$ regions are another pair. Moreover, the $k_y$ coordinates of Weyl points increase faster than those of the $k_x$ vector, and finally evolving to  ($\pm0.42\pi/h$, 0, $\pi/h$) and (0, $\pm0.98\pi/h$, $\pi/h$) at $\theta=90^\circ$, as denoted by the red and pink hollow circles in Fig. 2(b). It should be stressed that the frequencies of Weyl points here are not at the same frequency, and the tilted dispersions for each pair of type-II Weyl points are in different directions.

To show that the type-II Weyl points indeed exist in such a twisted layer-stacked photonic crystal, the 3D band structure for $\theta=30^\circ$ at $k_z=\pi/h$ plane is choosen to display the degenerate points and tilted cone dispersion, as presented in Fig. 2(c). We can see that the doubly degenerate bands are splitted when the twisting between different layers happens, with two nondegenerate bands separated into the high and low frequencies, respectively. Meanwhile, two middle bands degenerate at four Weyl points ($W$1, $W$2, $W$3 and $W$4) in momentum space with their locations in $(k_x, k_y, k_z)=(\pm0.86, \pm0.23, 1)\pi/h$, and  having the same frequency of 0.48$c/h$. Besides, it is clearly shown that the cone spectrum near these Weyl points is strongly tilted for $k_x$ direction. On the right side of Fig. 2(c), we further plot the linear degeneracy in the vicinity of these type-II Weyl points along $k_x$, $k_y$ and $k_z$ directions for fixed $k_y=0.23\pi/h$, $k_x=0.86\pi/h$ and $(k_x, k_y)=(0.86, 0.23)\pi/h$, respectively. For the $k_x$ direction, it is shown that the sign of group velocities for the two intersecting bands are the same around each degenerate point, which coincides with the features of type-II Weyl points. While for the other two directions, linear dispersions are also clearly presented. Therefore, combining with the characteristic of iso-frequency (at 0.48$c/h$) for these Weyl points, this twisted layer-stacked photonic system can actually exist four ideal type-II Wely points by tuning the twist angle $\theta$ within an appropriate range.

According to the bulk-edge correspondence, topological surface states should exist at the interface between a photonic type-II Weyl semimetal and the trivial material. To investigate the gapless surface states, we establish a supercell with 20 unit cells along $z$ direction and keep it infinite along the $x$ and $y$ directions, thus forming two truncated surfaces in a direction perpendicular to the $z$ axis, as shown in Fig. 3(c). Here we use air as the topological trivial material, and study the projected band structures corresponding to the case (Fig. 2(c)) at $\theta=30^\circ$. A full wave numerical calculation is adopted to calculate the band structures and eigenfields. For $k_x=0.86\pi/h$ (see Fig. 3(a)), the $W$2 and $W$4 Weyl points of opposite charges are projected onto the $xy$ surface, and there are two surface states (the red and blue curves) existed in the regions when $\left| k_y\right|\geq 0.23\pi/h$ (for the other pair of $W$1 and $W$3 Weyl points of $k_x=-0.86\pi/h$, similar results can also be found), indicating the nontrivial topology of these type-II Weyl points. To further verify the topological properties when $\left| k_y\right|\geq 0.23\pi/h$, we show the projected bands for $k_y=\pi/h$ and the eigenfields of surface states in Fig. 3(b) and 3(c). We find that two gapless surface states are located at an imcomplete bandgap, and they have the same sign of group velocity. This phenomenan is actually resulted from the tilted dispersions around type-II Weyl point. The |$\bf{E}$| of these two surface states shows that the eigenfields for the blue band are localized on the boundary in $-z$ direction, while the eigenfields for the red band are localized on the boundary in $+z$ direction, demonstrating the $\pm1$ charges of $W$2 and $W$4 Weyl points and the existence of topological surface states.  

\begin{figure*}[bth]
	\centering
	\includegraphics[height=8cm]{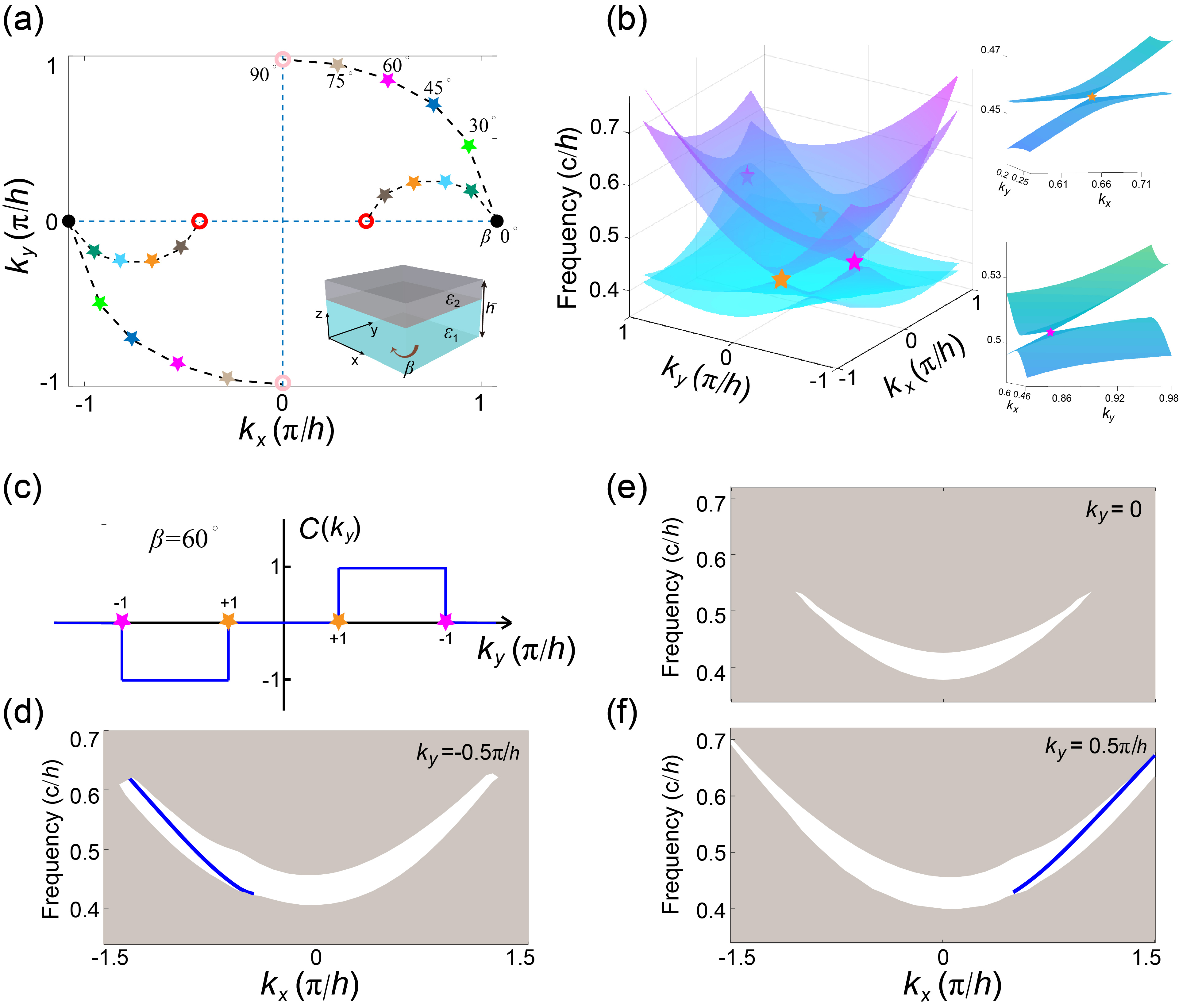}
	\caption{ (a) The evolutions of Dirac and Weyl points with twist angle $\beta$ from $0^\circ$ to $90^\circ$ for a structure where its unit cell is replaced by the bilayer anistropic material with different thicknesses (see the inset), $h_{\epsilon1}$=$2h/3$ and $h_{\epsilon2}$=$h/3$. (b) 3D band structure on the $k_x$-$k_y$ plane for $\beta=60^\circ$, the orange [($-0.66$, $-0.24$, 1)$\pi/h$, ($0.66$, $0.24$, 1)$\pi/h$]  and magenta [($-0.53$, $-0.86$, 1)$\pi/h$, ($0.53$, $0.86$, 1)$\pi/h$] stars mark four Weyl points, which are non-ideal at different frequencies. (c) Chern numbers and topological charges of the associated Weyl points in (b) as a function of $k_y$ for $\beta=60^\circ$. (d)-(f) The projected spectra at  $k_y$=-0.5$\pi/h$, $k_y$=0 and $k_y$=0.5$\pi/h$, respectively, where the gapless surface state( the blue lines) with single group velocity is observed when $k_y=\pm0.5\pi/h$.}
	\label{fg4}
\end{figure*}

 \section{Type-II Weyl points in twisted photonic crystals with bilayer unit cell} 

At last, we continue to explore the evolutions of location and shaping of the type-II Weyl points in momentum space by varying the permittivity distribution of unit cell structure to reduce its symmetry. As displayed in the inset of Fig. 4(a), the unit cell of twisted photonic crystal is replaced by the bilayer anisotropic materials $\epsilon_1$ and $\epsilon_2$ with different thicknesses of $h_{\epsilon1}$=$2h/3$ and $h_{\epsilon2}$=$h/3$, where the lower layer (dielectric $\epsilon_1$) has a twist angle $\beta$ respect to the $x$ axis. Here, $\epsilon_1$ and $\epsilon_2$ are still taken from Eq. 1, in addition to using $\beta$ to distinguish the twist angle. We can see that the parameter symmetry is thus broken in such a unit cell constructed of bilayer anisotropic materials. In Fig. 4(a), we show the evolution of Weyl points formed by the  degenerate bands (see the band structure in Fig. 4(b) for $\beta=60^\circ$ ) at $k_z=\pi/h$ plane when $\beta$ changes from $0^\circ$ to $90^\circ$. Starting from the initial case of $\beta=0^\circ$, it is shown that the type-II Dirac points (i.e., the black solid dots in FIG. 1(c)) are splitted into two pairs of type-II Weyl points, with each pair situated in the same quadrant, which is totally different from the locations in Fig.2. Besides, due to the fewer symmetry of parameters, the type-II Weyl points of each pair do not locate at the same frequency, implying the non-ideal type-II Weyl points. The evolutionary paths fitted in Fig. 4(a) further reveal that there are two paths for each pair of Weyl points with different features. Taking the Weyl points in the first quadrant as an example. One path has a shorter distance, with the $k_x$ coordinates of Weyl point decreases gradually to $0.42\pi/h$, while $k_y$ increases first and then decreases to 0 (the red hollow circle). Another path has a longer evolutionary distance in momentum space, where the $k_x$ coordinate declines from $1.07\pi/h$ to 0 and the $k_y$ coordinate increase from 0 to $0.98\pi/h$ (the pink hollow circle), respectively. It is shown that the final state corresponds exactly to the positions of Weyl points in Fig. 2(b) when $\theta=90^\circ$, which can be easily understood that the twisted structure in Fig. 1(a) is reduced to the asymmetrical case in Fig. 4(a) when $\theta=90^\circ$. The results are echoed back and forth.

In Fig. 4(b), we show the 3D band structure for $\beta=60^\circ$ at $k_z=\pi/h$ plane and mark the locations of four non-ideal type-II Weyl point by the orange and magenta stars. It is intuitive to see that each pair of Weyl points are located in the same quadrant with different frequencies, and the locations of these Weyl points in momentum space are ($-0.66$, $-0.24$, 1)$\pi/h$ and ($0.66$, $0.24$, 1)$\pi/h$ [the orange stars], ($-0.53$, $-0.86$, 1)$\pi/h$ and ($0.53$, $0.86$, 1)$\pi/h$ [the magenta stars], respectively. In addition, linear dispersions around these Weyl points can also be observed, including the tilted cone dispersion in $k_x$ direction for the two Weyl points marked as orange stars, while for the other two Weyl points (magenta stars), the linear dispersion is tilted in $k_y$ direction, as denoted by the inset. This unique tilting structure of each pair of Weyl points is different from that in Fig. 2. Next we will study the projected band structures of such twisted photonic crystal at different $k_y$ to investigate the existence of surface states. The topological charges for Weyl points are $\pm1$ as shown in Fig. 4(c), where the topological transitions take place, and the positions and charges of Weyl points thus induce the $k_y$ dependence of Chern number[16]. Three representative $k_y$ values lying between the Weyl points with different charges are taken to study the band structure projected along the $z$ direction (i.e., the Weyl photonic
crystal is truncated in $z$ direction ). When $k_y=-0.5\pi/h$, the calculated dispersions near the frequencies of Weyl points are displayed in Fig. 4(d), where the surface state is plotted in blue curve and the bulk states are shown in grey. It can be seen that there exists a gapless surface state with negative group velocity, which is consistent with the Chern number $C=-1$ in Fig. 4(c) and is resulted from the opposite topological charges of type-II Weyl points. As $k_y$ increases to $0$ (see Fig. 4(e)), there is no surface state located in the projected bandgap due to the trivial phase between two Weyl points with the same charge. Next, as $k_y$ further increases to $0.5\pi/h$, the projected band structure can be found in Fig. 4(f), where the  surface state reappears with positive group velocity, corresponding to the +1 Chern number in Fig. 4(c). The surface states thus verify the topology of these four non-ideal type-II Weyl points. 

Finally, we propose that the naturally anisotropic materials with elliptic in-plane dispersion may be used to realize the ideal model of such a layer-stacked 1D photonic crystals, though losses can be introduced due to the dispersion of practical material[47,48]. By controlling the twist angle between layers, the type-II Weyl points residing in different positions can be found in momentum space.

\section{Conclusion and outlook}
To summarize, we theoretically designed the layer-stacked twisted photonic crystals using anisotropic dielectrics, and proved the existence of photonic type-II Weyl points in such a system. By symmetrically tuning the twist angle in a unit cell consisting of trilayer dielectrics, we split two type-II Dirac points into the four ideal type-II Weyl points, and analyzed the linear dispersions around them. The evolutions of these Weyl points with respect to twist angle is detailly studied in a rotation period. In addition, gapless surface states are found at the truncated surface of such photonic crystal slab. We further explore a twisted structure with fewer symmetry and obtain the non-ideal type-II Weyl points, verifying the existence of surface states with single group velocity. In the twisted photonic system, naturally anisotropic materials are expected to be used as a platform for studying the topological physics of Weyl points, or other semimetal phases, which do not require the complex structural design as in 3D photonic crystals and metamaterials, and may open up new ideas for exploring topological states in systems composed of naturally anisotropic materials.

\section*{Acknowledgements}
This work was supported by the National Natural Science Foundation of China under the Grant nos. 11874311/11904060; Huaqiao University (605-50Y21003);  
the Fundamental Research Funds for the Central Universities (Grant No. 20720200074).

\section*{References}

[1] N. P. Armitage, E. J. Mele, and A. Vishwanath, Weyl and dirac semimetals in three dimensional solids, Reviews of Modern Physics 90, 015001 (2018)
\vspace{1ex} 

\noindent{[2] X. G. Wan, A. M. Turner, A. Vishwanath, et al., Topological semimetal and fermi-arc surface
states in the electronic structure of pyrochlore iridates, Physical Review B 83, 205101 (2011).}
\vspace{1ex} 

\noindent{[3] S. Y. Xu, I. Belopolski, N. Alidoust, et al., Discovery of a weyl fermion semimetal and
topological fermi arcs, Science 349, 613–617 (2015).} 
\vspace{1ex} 

\noindent{[4] B. Q. Lv, H. M. Weng, B. B. Fu, et al., Experimental discovery of weyl semimetal taas,
Physical Review X 5, 031013 (2015).}
\vspace{1ex} 

\noindent{[5] S. M. Young, S. Zaheer, J. C. Y. Teo, et al., Dirac semimetal in three dimensions, Physical
Review Letters 108, 140405 (2012).}
\vspace{1ex} 

\noindent{[6] Z. K. Liu, B. Zhou, Y. Zhang, et al., Discovery of a three-dimensional topological dirac
semimetal, na3bi, Science 343, 864–867 (2014).}
\vspace{1ex} 

\noindent{[7] A. A. Burkov, M. D. Hook, and L. Balents, Topological nodal semimetals, Physical Review B 84, 235126 (2011).}
\vspace{1ex} 

\noindent{[8] R. Bi, Z. B. Yan, L. Lu, et al., “Nodal-knot semimetals,” Physical Review B 96, 201305 (2017).}
\vspace{1ex}

\noindent{[9] J.-H. Jiang, “Tunable topological weyl semimetal from simple-cubic lattices with staggered
fluxes,” Physical Review A 85(3), 033640 (2012).}
\vspace{1ex}

\noindent{[10] A. A. Soluyanov, D. Gresch, Z. J. Wang, et al., “Type-ii weyl semimetals,” Nature 527, 495–498 (2015).}
\vspace{1ex}

\noindent{[11] B. H. Yan and C. Felser, Topological Materials: Weyl Semimetals, vol. 8, 337–354 (2017).}
\vspace{1ex}

\noindent{[12] L. Lu, L. Fu, J. D. Joannopoulos, et al., “Weyl points and line nodes in gyroid photonic
crystals,” Nature Photonics 7, 294–299 (2013).}
\vspace{1ex}

\noindent{[13] L. Lu, Z. Y. Wang, D. X. Ye, et al., “Experimental observation of weyl points,” Science 349, 622–624 (2015).}
\vspace{1ex}

\noindent{[14] W. J. Chen, M. Xiao, and C. T. Chan, “Photonic crystals possessing multiple weyl points
and the experimental observation of robust surface states,” Nature Communications 7, 13038
(2016).}
\vspace{1ex}

\noindent{[15] W. L. Gao, B. Yang, M. Lawrence, et al., “Photonic weyl degeneracies in magnetized
plasma,” Nature Communications 7, 12435 (2016).}
\vspace{1ex}

\noindent{[16] H. X. Wang, L. Xu, H. Y. Chen, et al., “Three-dimensional photonic dirac points stabilized
by point group symmetry,” Physical Review B 93(23), 235155 (2016).}
\vspace{1ex}

\noindent{[17] Z. Yang, M. Xiao, F. Gao, et al., “Weyl points in a magnetic tetrahedral photonic crystal,”
Optics Express 25, 15772–15777 (2017).}
\vspace{1ex}

\noindent{[18] H. X. Wang, Y. Chen, Z. H. Hang, et al., “Type-ii dirac photons,” Npj Quantum Materials 2, 54 (2017).}
\vspace{1ex}

\noindent{[19] J. Noh, S. Huang, D. Leykam, et al., “Experimental observation of optical weyl points and
fermi arc-like surface states,” Nature Physics 13, 611–617 (2017).}
\vspace{1ex}

\noindent{[20] B. A. Yang, Q. H. Guo, B. Tremain, et al., “Direct observation of topological surface-state
arcs in photonic metamaterials,” Nature Communications 8, 7 (2017).}
\vspace{1ex}

\noindent{[21] W. L. Gao, B. Yang, B. Tremain, et al., “Experimental observation of photonic nodal line
degeneracies in metacrystals,” Nature Communications 9, 950 (2018).}
\vspace{1ex}

\noindent{[22] Q. H. Yan, R. J. Liu, Z. B. Yan, et al., “Experimental discovery of nodal chains,” Nature
Physics 14, 461–464 (2018).}
\vspace{1ex}

\noindent{[23] Z. F. Xiong, R. Y. Zhang, R. Yu, et al., “Hidden-symmetry-enforced nexus points of nodal
lines in layer-stacked dielectric photonic crystals,” Light-Science and Applications 9(1), 176 (2020).}
\vspace{1ex}

\noindent{[24] H. F. Wang, S. K. Gupta, B. Y. Xie, et al., “Topological photonic crystals: a review,” Frontiers
of Optoelectronics 13, 50–72 (2020).}
\vspace{1ex}

\noindent{[25] M. Kim, Z. Jacob, and J. Rho, “Recent advances in 2d, 3d and higher-order topological
photonics,” Light-Science and Applications 9(1), 130 (2020).}
\vspace{1ex}

\noindent{[26] M. Xiao, W. J. Chen, W. Y. He, et al., “Synthetic gauge flux and weyl points in acoustic
systems,” Nature Physics 11, 920–924 (2015).}
\vspace{1ex}

\noindent{[27] H. Ge, X. Ni, Y. Tian, et al., “Experimental observation of acoustic weyl points and topologial surface states,” Physical Review Applied 10, 014017 (2018).}
\vspace{1ex}

\noindent{[28] H. L. He, C. Y. Qiu, L. P. Ye, et al., “Topological negative refraction of surface acoustic waves
in a weyl phononic crystal,” Nature 560, 61–+ (2018).}
\vspace{1ex}

\noindent{[29] W. Y. Deng, J. Y. Lu, F. Li, et al., “Nodal rings and drumhead surface states in phononic
crystals,” Nature Communications 10, 1769 (2019).}
\vspace{1ex}

\noindent{[30] M. Xiao, X. Q. Su, and S. H. Fan, “Nodal chain semimetal in geometrically frustrated systems,” Physical Review B 99, 094206 (2019).}
\vspace{1ex}

\noindent{[31] B. Y. Xie, H. Liu, H. Cheng, et al., “Dirac points and the transition towards weyl points in
three dimensional sonic crystals,” Light-Science and Applications 9, 201 (2020).}
\vspace{1ex}

\noindent{[32] Y. H. Lu, N. Y. Jia, L. Su, et al., “Probing the berry curvature and fermi arcs of a weyl circuit,”
Physical Review B 99, 020302 (2019).}
\vspace{1ex}

\noindent{[33] M. Ezawa, “Magnetic second-order topological insulators and semimetals,” Physical Review
B 97, 155305 (2018).}
\vspace{1ex}

\noindent{[34] Z.-X. Li, Y. Cao, and P. Yan, “Topological insulators and semimetals in classical magnetic
systems,” Physics Reports (2021).}
\vspace{1ex}

\noindent{[35] H. X. Wang, Z. K. Lin, B. Jiang, et al., “Higher-order weyl semimetals,” Physical Review
Letters 125, 146401 (2020).}
\vspace{1ex}

\noindent{[36] S. A. A. Ghorashi, T. H. Li, and T. L. Hughes, “Higher-order weyl semimetals,” Physical
Review Letters 125, 266804 (2020).}
\vspace{1ex}

\noindent{[37] Q. Wei, X. Zhang, W. Deng, et al., “Higher-order topological semimetal in acoustic crystals,” Nature Materials , 1–6 (2021).}
\vspace{1ex}

\noindent{[38] L. Luo, H.-X. Wang, Z.-K. Lin, et al., “Observation of a phononic higher-order weyl semimetal,” Nature Materials, 20, 794-799 (2021).}
\vspace{1ex}

\noindent{[39] B. Xie, H. Liu, H. Cheng, et al., “Experimental realization of type-ii weyl points and fermi
arcs in phononic crystal,” Physical review letters 122, 104302 (2019).}
\vspace{1ex}

\noindent{[40] Z. Yang and B. Zhang, “Acoustic type-ii weyl nodes from stacking dimerized chains,” Physical review letters 117, 224301 (2016).}
\vspace{1ex}

\noindent{[41] R. Li, B. Lv, H. Tao, et al., “Ideal type-ii weyl points in topological circuits,” arXiv preprint
arXiv:1910.03503 (2019).}
\vspace{1ex}

\noindent{[42] B. Yang, Q. Guo, B. Tremain, et al., “Ideal weyl points and helicoid surface states in artificial
photonic crystal structures,” Science 359, 1013–1016 (2018).}
\vspace{1ex}

\noindent{[43] X. Huang, W. Deng, F. Li, et al., “Ideal type-ii weyl phase and topological transition in
phononic crystals,” Physical Review Letters 124, 206802 (2020).}
\vspace{1ex}

\noindent{[44] Y. Yang, Z. Gao, X. Feng, et al., “Ideal unconventional weyl point in a chiral photonic metamaterial,” Physical Review Letters 125, 143001 (2020).}
\vspace{1ex}

\noindent{[45] M. Li, J. Song, and Y. Jiang, “Photonic topological weyl degeneracies and ideal type-i weyl
points in the gyromagnetic metamaterials,” Physical Review B 103(4), 045307 (2021).}
\vspace{1ex}

\noindent{[46] M.-L. Chang, M. Xiao, W.-J. Chen, et al., “Multiple weyl points and the sign change of
their topological charges in woodpile photonic crystals,” Physical Review B 95(12), 125136
(2017).}
\vspace{1ex}

\noindent{[47] W. L. Ma, P. Alonso-Gonzalez, S. J. Li, et al., “In-plane anisotropic and ultra-low-loss polaritons in a natural van der waals crystal,” Nature 562, 557 (2018).}
\vspace{1ex}

\noindent{[48] G. Hu, Q. Ou, G. Si, et al., “Topological polaritons and photonic magic angles in twisted $\alpha$-MoO$_3$ bilayers,” Nature 582, 209–213 (2020).}
\vspace{1ex}

\end{document}